\documentclass[a4paper,british,oldversion]{aa}
\usepackage[T1]{fontenc}
\usepackage[latin9]{inputenc}
\usepackage{color}
\usepackage{amsmath}
\usepackage{graphicx}
\usepackage{amssymb}
\usepackage[authoryear]{natbib}

\makeatletter

\usepackage{xcolor}
\usepackage{ifthen}
\authorrunning{R.G. Izzard et al.}
\titlerunning{Ba Stars and WD Kicks}
 \date{}

\newcommand{\Change}[1]{#1}

\newcommand{\Pcirc}{4,\!000}
\newcommand{\Pmax}{10^{4}}
\newcommand{\Pgap}{1,\!000}
\usepackage{amsbsy} 

\makeatother

\usepackage{babel}

\begin{document}

\title{White-Dwarf Kicks and Implications for Barium Stars}

\author{Robert G. Izzard\inst{1,2,4}, Tyl Dermine\inst{1} \and Ross P.
Church\inst{3,4} }

\institute{Institut d'Astronomie et d'Astrophysique, Université Libre de Bruxelles,
Boulevard du Triomphe, B-1050 Brussels, Belgium \and \Change{Argelander-Institut
für Astronomie, University of Bonn, Auf dem Hügel 71, D-53121 Bonn,
Germany%
\thanks{\Change{Present address.}%
}}\and Lund Observatory, Box 43, SE-221 00, Lund, Sweden \and Centre
for Stellar and Planetary Astrophysics, School of Mathematical Sciences,
Monash University, VIC 3800, Australia}

\abstract{The formation mechanism of the barium stars is thought to be well
understood. Barium-rich material, lost in a stellar wind from a thermally-pulsing
asymptotic-giant branch star in a binary system, is accreted by its
companion main-sequence star. Now, many millions of years later, the
primary is an unseen white dwarf and the secondary has itself evolved
into a giant which displays absorption lines of barium in its spectrum
and is what we call a barium star. A similar wind-accretion mechanism
is also thought to form the low-metallicity CH and carbon-enhanced
metal-poor stars. Qualitatively the picture seems clear but quantitatively
it is decidedly murky: several key outstanding problems remain which
challenge our basic understanding of binary-star physics. Barium stars
with orbital periods less than about $\Pcirc\,\mathrm{days}$ should
-- according to theory -- be in circular orbits because of tidal dissipation,
yet they are often observed to be eccentric. Only one barium-star
period longer than $\Pmax\,\mathrm{days}$ has been published although
such stars are predicted to exist in large numbers. In this paper
we attempt to shed light on these problems. First, we consider the
impact of kicking the white dwarf at its birth, a notion which is
supported by independent evidence from studies of globular clusters.
Second, we increase the amount of orbital angular momentum loss during
wind mass transfer, which shrinks barium-star binaries to the required
period range. We conclude with a discussion of possible physical mechanisms
and implications of a kick, such as the break up of wide barium-star
binaries and the limits imposed on our models by observations.}

\maketitle

\section{Introduction}

The barium stars are population I red giants which show strong absorption
lines of barium in their spectra. They make up about 1\% of all G/K
giants. They are not evolved enough to have produced the barium themselves,
rather it was made in a companion star which has long since ceased
nuclear burning. This is borne out by the observation that all barium
stars are in binary stellar systems \citep{1990mcclure_woodsworth}.

Barium is an $s$-process element made during the thermally-pulsing
asymptotic giant branch (TPAGB) stellar evolutionary phase. In sufficiently
close binaries containing a TPAGB star, the primary becomes large
enough that Roche-lobe overflow (RLOF) terminates the TPAGB and lessens
the production of barium, preventing barium-star formation. At large
separations the strong stellar wind of the TPAGB star leads to mass
accretion on the secondary and this is the favoured channel for barium
star formation. An intermediate regime may exist in which some wind
mass-transfer occurs before RLOF begins (e.g. \citealp{1995MNRAS.277.1443H}).

While the wind-accretion scenario is generally accepted, it does not
explain the distribution of eccentricities and periods of barium stars.
Most barium stars with $P\lesssim500\,\mathrm{days}$ are in (near-)circular
orbits as predicted by tidal circularisation theory \citep[e.g.][]{1977A+A57-383Z,1989A&A...220..112Z}.
Most of the rest of the barium stars have periods between $500$ and
$10^{4}\,\mathrm{days}$ and eccentricities $0\leq e\lesssim0.4$
\citep{1998A&A...332..877J}.

Population synthesis studies remain the best way to study the orbital
characteristics of barium stars in a statistical manner. Models such
as those of \citet{2003ASPC..303..290} confirm that our understanding
of barium-star formation is incomplete. While the models confirm that
short-period systems are indeed (almost) circular, they predict circularisation
for all systems with periods shorter than about $\Pcirc\,\mathrm{days}$,
which is not seen in the observations. They also predict a long-period
tail of eccentric barium stars extending to $10^{5}\,\mathrm{days}$,
also not seen in the observations. While binaries with periods greater
than $10^{4}\,\mathrm{days}$ are difficult to detect, the fraction
of barium stars with measured periods is large ($35/37$ for strong
barium stars) so a population with undetected, longer-period orbits
can be ruled out as an explanation of the discrepancy between models
and observations. The combination of large eccentricities and relatively
short periods of the barium stars remains an unresolved problem.

If tides are as efficient as predicted a mechanism must exist which
generates eccentricity in \Change{barium star systems at the end
of the TPAGB phase of the primary.} This is evident from the observed
eccentricities of \Change{binary} post-AGB stars which show a distribution
strikingly similar to the barium stars. Many sources of eccentricity
have been investigated such as enhanced mass-transfer at periastron
\citep{2000A&A...357..557S}, interaction with a circumbinary disc
\citep{2007BaltA..16..104F} and eccentricity pumping induced by a
wind-RLOF hybrid mass transfer \citep{2008A&A...480..797B}.

Recently and independently of the study of barium stars several authors
have suggested that white dwarfs are kicked when they are born, i.e.
at the end of the TPAGB phase. This result is based on a comparison
of the velocity dispersion of young white dwarfs to old white dwarfs
and main-sequence stars in globular clusters \citep{2008MNRAS.383L..20D}.
The young white dwarfs have a higher velocity dispersion than expected
by a few $\mathrm{km}\,\mathrm{s}^{-1}$ and are cited as a possible
mechanism to prevent globular cluster collapse \citep{2008MNRAS.385..231H,2009ApJ...695L..20F}.
If white-dwarf kicks occur they must affect the orbital parameters
of the barium stars as well as their low-metallicity equivalents,
the carbon-enhanced metal-poor (CEMP) and CH stars.

\Change{The cause of the white-dwarf kicks is unknown. Candidate
mechanisms include asymmetric mass loss during the AGB \citep[e.g.][]{2003ApJ...595L..53F},
magnetic fields \citep{1998A&A...333..603S} and perhaps  non-radial
stellar pulsations. Dynamical interactions inside the globular cluster
may also play a role but it is not clear why these should affect only
young white dwarfs. Studies of planetary nebulae may be able to constrain
any connection between asymmetric mass loss \citep[e.g.][]{1998AJ....116.1357S},
magnetic fields \citep{2007AJ....133..987L} and a kick to the central,
proto-white dwarf, star. However, it is often difficult to determine
the velocity of the central star compared to the expanding planetary
nebula because of interaction of the nebula with a non-uniform interstellar
medium.}

In this paper we investigate the possibility that a kick to the newly-born
white dwarf is the cause of the eccentricity in the barium stars.
In Section~\ref{sec:models} we describe our population synthesis
models, in Section~\ref{sec:results} we present our results and
compare them to observed barium star periods and eccentricities, Section~\ref{sec:discussion}
discusses successes and potential problems of our model before we
conclude with Section~\ref{sec:conclusions}.

\section{Modelling Ba stars with white-dwarf kicks}

\label{sec:models}Our binary population synthesis model is based
on that of \citet*{2002MNRAS_329_897H} with nucleosynthesis as described
by \citet{Izzard_et_al_2003b_AGBs,2006A&A...460..565I,2009A&A...508.1359I}.
Binary systems have initial primary masses $M_{1}$ distributed according
to the \citet*{KTG1993MNRAS-262-545K} initial mass function between
$1.2$ and $3\,\mathrm{M}_{\odot}$, a flat distribution in the mass
ratio $q=M_{1}/M_{2}$ (where $M_{1}\geq M_{2}$) and a flat distribution
in $\log a$ between $400$ and $10^{5}\,\mathrm{R}_{\odot}$. Initial
eccentricities are chosen from a distribution $f(e)\propto e$ between
$0$ and $1$. We assume that all stars form in binaries and our models
here have a grid resolution $N_{\mathrm{M1}}\times N_{\mathrm{M2}}\times N_{a}\times N_{e}=30\times30\times60\times30$.

We model tides following \citet*{2002MNRAS_329_897H} who base their
model on the formalisms of \citet{1977A+A57-383Z} and \citet{1981A+A....99..126H}.
The tidal circularisation timescale during the TPAGB $\tau_{\mathrm{circ}}\sim10^{4}(a/3R)^{8}\,\mathrm{years}$
(e.g. \citealp{2000A&A...357..557S}). This implies that tides rapidly
circularise any binary containing an AGB star with a separation less
than a few stellar radii.

Systems which are close enough to enter Roche-lobe overflow during
the AGB undergo common envelope evolution with rapid circularisation,
ejection of the stellar envelope and little barium production and
accretion onto the secondary. These systems lead to either little
barium production or perhaps short-period barium stars. Some authors
have speculated about stable case-C mass transfer which may lead to
the formation of the shortest-period barium stars \citep[e.g.][]{1995MNRAS.277.1443H}
and the intriguing possibility of a radiation-distorted Roche geometry
\citep{2009A&A...507..891D}. We do not explore this channel in detail
here because we focus on the wind mass-transfer scenario.

Our nucleosynthesis model includes third dredge up with an efficiency
given by \citet{Parameterising_3DUP_Karakas_Lattanzio_Pols} and $s$-process
abundances based on the models of \citet{2001ApJ...557..802B}. Our
initial metallicity is $Z=0.008$ with an abundance mixture according
to \citet{1989GeCoA..53..197A}. \Change{The $^{13}\mathrm{C}$ pocket
efficiency, $^{13}\xi$, is a multiplicative factor used to change
the amount of of $^{13}\mathrm{C}$ relative to the $2.8\times 10^{-6}\,\mathrm{M}_{\odot}$
in the standard pocket of \citet{1998ApJ...497..388G}. We set it}
to $1$ in order to make sufficient barium stars at $Z=0.008$ (about
$1\%$ of GK giants). At $Z=0.02$ even with $^{13}\xi=2$ we cannot
make enough Ba stars, however the amount of third dredge up is also
rather uncertain and can be increased artificially to obtain the desired
$1\%$ ratio (as noted by \citealp{1995MNRAS.277.1443H}). We do not
try to constrain either $^{13}\xi$ or the amount of third dredge
up, this has been attempted by others (e.g. \citealp{2006MmSAI..77..879B},
see also the discussion in Section~\ref{sub:mild-strong}).

Mass loss during the TPAGB is parameterised by the formula of \citet{1993ApJ...413..641V}
and mass accretion onto the secondary is at the rate given by \citet{1944MNRAS.104..273B}
as described in \citet*[Eq. 6]{2002MNRAS_329_897H} with an efficiency
parameter $\alpha_{\mathrm{W}}=1.5$ and an accretion rate limited
by $\left|\dot{M}_{2}\right|<0.8\left|\dot{M}_{1}\right|$ (where
star~$1$ is the donor, star~$2$ the accretor). At the end of the
TPAGB phase of the primary star a kick is applied to the nascent white
dwarf with a fixed speed $\sigma_{\mathrm{WD}}$ and random direction
according to the method described in \citet*[appendix A1, but without the Maxwellian velocity distribution]{2002MNRAS_329_897H}.
We apply the kick whether the TPAGB is terminated by a stellar wind
or common-envelope ejection.

We count stars as giants after they leave the main sequence \Change{(as
defined by \citealp*{2002MNRAS_329_897H})} and before they become
white dwarfs. \Change{ We select stars with only G and K spectral
types corresponding to $3620\leq T_{\mathrm{eff}}/\mathrm{K}\leq5150$
\citep{Jaschek_Jaschek}.} Of these, strong barium stars have $[\mathrm{Ba}/\mathrm{Fe}]\geq0.5$
and mild barium stars have $0.2\leq[\mathrm{Ba}/\mathrm{Fe}]<0.5$.

\section{Simulated Ba-star populations}

\label{sec:results}In this section we present the results of our
population syntheses with the systematic inclusion of progressively
speculative physics as required to reproduce the observed $e-\log P$
distribution of the barium stars.

We compare our models with the observations of \citet{1998A&A...332..877J}
but without the triple system BD~+38$^{\circ}$~118. For the strong
barium stars the sample is almost complete (35 out of 37 have measured
periods and eccentricities). Our weak barium star sample is augmented
with observations of 56~Uma \citep{2008Obs...128..176G} and the
latest results of the \emph{HERMES} survey (Dermine and Jorissen,
private communication). Very few barium stars do not have a measured
period and eccentricity and the missing stars do not affect any of
the general conclusions we present in this paper.

\subsection{Our canonical model}

\label{sub:canonical}%
\begin{figure}
\includegraphics[bb=50bp 100bp 500bp 770bp,angle=270,scale=0.42]{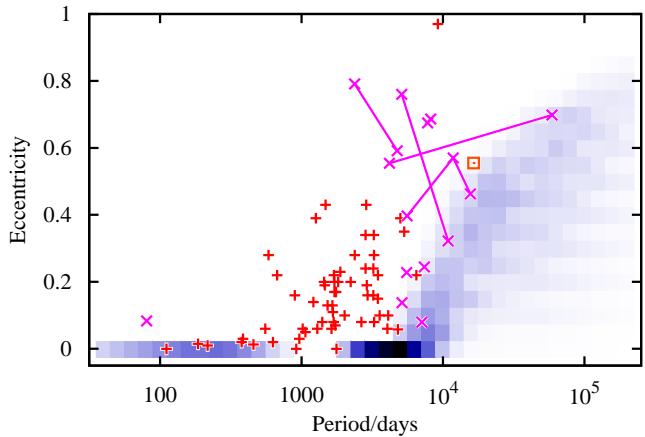}

\caption{\label{fig:canonical-ba-elogP}Our canonical barium-star $e-\log P$
distribution. Darker greyscale/colour indicates a greater number density
of systems. The plus symbols \textcolor{red}{$\pmb{+}$} are the observations
of \citet{1998A&A...332..877J} without the triple system BD~+38$^{\circ}$118~,
the diagonal crosses \textcolor{magenta}{$\pmb{\times}$} are the
latest \emph{HERMES} observations (Dermine and Jorissen, private communication)
and the square \textcolor{red}{\color{orange}{$\pmb{\square}$}} is
56~Uma \citep{2008Obs...128..176G}. Where points are joined by lines
there are multiple possible orbital solutions.}

\end{figure}
Our canonical model result, i.e. without white-dwarf kicks or otherwise
modified physics, is shown in Fig.~\ref{fig:canonical-ba-elogP}
and well reproduces previous studies such as \citet{2003ASPC..303..290}
and \citet{2008A&A...480..797B}. Our simulated barium-star systems
with periods of less than about $\Pcirc\,\mathrm{days}$ are circular.
At longer periods our modelled barium stars are eccentric and some
have periods as large as $10^{5}\,\mathrm{days}$. The observed barium
stars do not match either property of our simulated systems, most
being eccentric and having periods typically less than $10^{4}\,\mathrm{days}$.

An obvious solution to the long-period tail is to decrease the wind
accretion efficiency through the $\alpha_{\mathrm{W}}$ parameter.
However, this has the effect of reducing the number ratio of barium
to G/K giants to significantly below $1\%$ so is not justified by
the observations. It may be that a more complicated dependence of
$\alpha_{\mathrm{W}}$ on the orbital separation allows the formation
of barium stars only at periods less than $10^{4}\,\mathrm{days}$.
However, without a suitable formalism it is difficult to test anything
but an even more ad-hoc solution to this problem than simply altering
the constant $\alpha_{\mathrm{W}}$. As such, we keep $\alpha_{\mathrm{W}}=1.5$
in the remainder of this paper (except where stated otherwise).

The gap in our simulated population at $\sim\Pgap\,\mathrm{days}$
is because of our chosen common envelope evolution prescription. The
AGB core and companion star spiral together inside the common envelope
so the period shortens, in some cases to less than $100\,\mathrm{days}$.
Some observed barium stars indeed have such short periods, but are
not necessarily circular. We discuss our common envelope prescription
in more detail in Section~\ref{sub:Comenv} below.

\subsection{White-dwarf kicks}

\label{sub:with-kicks}%
\begin{figure}
\includegraphics[bb=50bp 100bp 500bp 770bp,angle=270,scale=0.42]{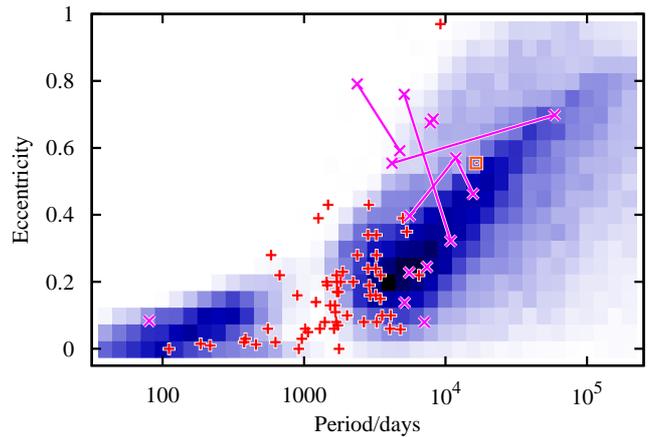}

\caption{\label{fig:ba-elogP-with-kicks}As Fig.~\ref{fig:canonical-ba-elogP}
but for a population including $4\,\mathrm{km}\,\mathrm{s}^{-1}$
natal white-dwarf kicks.}

\end{figure}
In Fig.~\ref{fig:ba-elogP-with-kicks} we show our simulated barium-star
population with the inclusion of a $4\,\mathrm{km}\,\mathrm{s}^{-1}$
kick at the moment of birth of the white dwarf. Our simulation and
the observations are in much better agreement than in the model of
the previous section (c.f. Fig.~\ref{fig:canonical-ba-elogP}). Even
the short-period systems acquire a small eccentricity. However, the
problematic long-period tail extends to even longer periods than in
our canonical model. These systems are weakly bound and a kick forces
the binary into a wider orbit. However, too strong a kick leads to
many broken binaries which are not observed, see Section~\ref{sub:disruption}.

\subsection{Common envelope evolution}

\label{sub:Comenv}%
\begin{figure}
\includegraphics[bb=50bp 100bp 500bp 770bp,angle=270,scale=0.42]{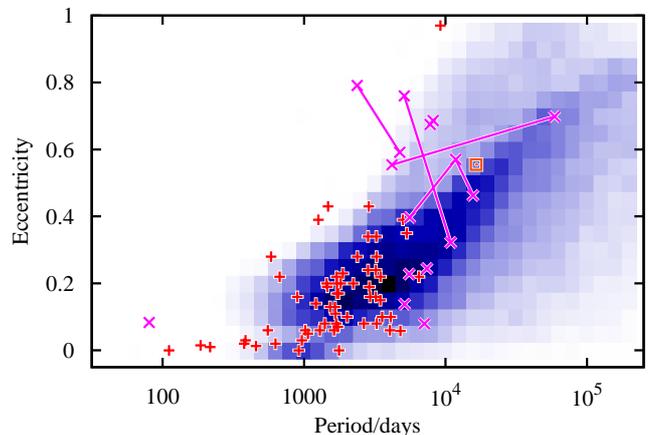}

\caption{\label{fig:ba-elogP-comenv}As Fig.~\ref{fig:ba-elogP-with-kicks}
but for a population with efficient common-envelope ejection ($\lambda_{\mathrm{ion}}=0.1$).}

\end{figure}
Our common-envelope prescription is based on the energy-balance formalism
of \citet{2002MNRAS_329_897H} with $\alpha=1$ and $\lambda$ fitted
to detailed stellar models \citep{2000A&A...360.1043D}. It leads
to the spiral in of systems with periods around $1,\!000\,\mathrm{days}$
and hence the period gap observed in Fig.~\ref{fig:canonical-ba-elogP}
(as noted by e.g. \citealp{1995MNRAS.277.1443H}). The period gap
is not seen in the observed barium-star period distribution and so
one is forced to ask how it can be removed. 

The most obvious solution is to increase the efficiency of common-envelope
ejection and hence reduce the amount of spiral-in before the envelope
is lost. We do this by adding a fraction $\lambda_{\mathrm{ion}}$
of the recombination energy of the envelope to the energy used to
eject the envelope. In the simulations described in Sections~\ref{sub:canonical}~and~\ref{sub:with-kicks},
$\lambda_{\mathrm{ion}}=0$. However, because TPAGB envelopes are
cool and weakly bound, even a small value of $\lambda_{\mathrm{ion}}$,
just a few per cent, is sufficient to close the period gap, as shown
in Fig.~\ref{fig:ba-elogP-comenv} for $\lambda_{\mathrm{ion}}=0.1$.
For the shortest-period barium stars a smaller value of $\lambda_{\mathrm{ion}}$
is required which suggests it may not take a constant value as we
have assumed. We note that use of the common-envelope prescription
of \citet{2005MNRAS.356..753N} gives a similar result.

\subsection{Orbital angular momentum}

\label{sub:Jorb}%
\begin{figure}
\includegraphics[bb=50bp 100bp 500bp 770bp,angle=270,scale=0.42]{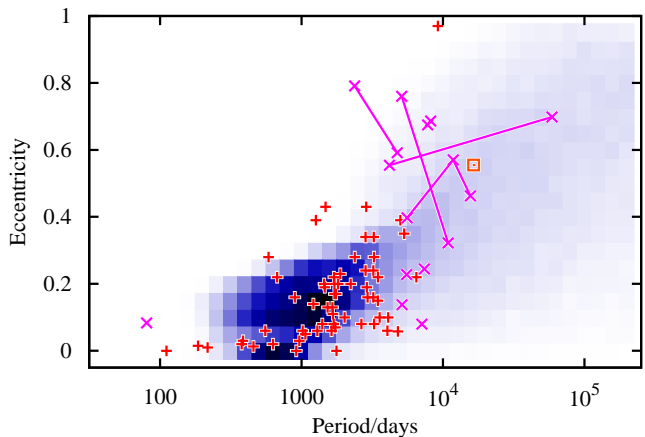}

\caption{\label{fig:ba-elogP-jorb}As Fig.~\ref{fig:ba-elogP-comenv} but
for a population very efficient orbital angular momentum loss according
to the \citet{2002MNRAS_329_897H} formalism (Eq.~\ref{eq:hurley-jorbdot}
and an accretion parameter $\alpha_{\mathrm{W}}=100$, see text for
details).}

\end{figure}
Wind accretion is not likely to be $100\%$ efficient, that is, some
material is always lost from the system. The material that is lost
carries away orbital angular momentum so the orbital separation changes.
It is not clear how much angular momentum is lost per unit mass ejected.
Our standard model, as presented in the sections above, uses the prescription
of \citet{2002MNRAS_329_897H} which provides the following formula,

\begin{eqnarray}
\dot{J}_{\mathrm{orb}} & = & \frac{J_{\mathrm{orb}}}{M_{1}+M_{2}}\left(\dot{M}_{1}\frac{M_{2}}{M_{1}}-\dot{M}_{2}\right)\,,\label{eq:hurley-jorbdot}\end{eqnarray}
which is \emph{always} \emph{negative} because $\dot{M}_{1}<0$ and
$\dot{M}_{2}>0$ (where star $1$ is the donor TPAGB star and star
$2$ is the accretor). When wind accretion is very efficient, the
term $-\dot{M}_{2}M_{1}$ is significant, the orbit shrinks and we
obtain the results shown in Fig.~\ref{fig:ba-elogP-jorb}. In the
\citet{2002MNRAS_329_897H} model efficient wind accretion can be
simulated with a large value of $\alpha_{\mathrm{W}}$\Change{. The
accretion rate is then limited to $\left|\dot{M}_{2}\right|<0.8\left|\dot{M}_{1}\right|$.}
Figure~\ref{fig:ba-elogP-jorb} shows the case with $\alpha_{\mathrm{W}}=100$
which is effectively $80\%$ efficient accretion because of the imposed
limit $\left|\dot{M}_{2}\right|\leq0.8\left|\dot{M}_{1}\right|$.
Most of the long-period barium stars in our model shrink their orbits
to periods less than $10^{4}\,\mathrm{days}$ so agree much better
with the observations.

\begin{figure}
\includegraphics[bb=50bp 100bp 500bp 770bp,angle=270,scale=0.42]{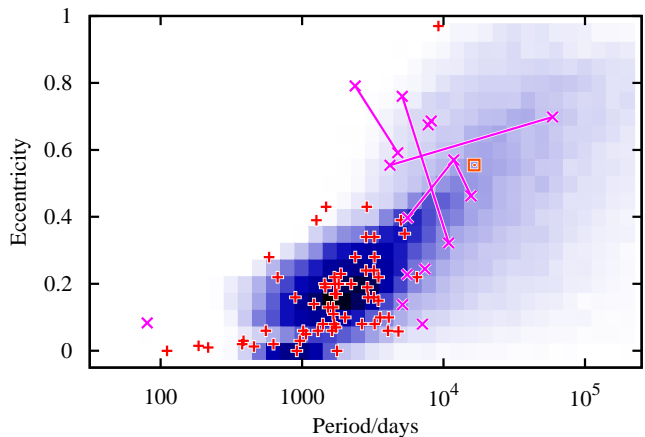}

\caption{\label{fig:ba-elogP-jorb-l-formalism}As Fig.~\ref{fig:ba-elogP-jorb}
but with our alternative orbital angular momentum loss formula (Eq.~\ref{eq:my-jorbdot}
with $l=2$ and an accretion parameter $\alpha_{\mathrm{W}}=1.5$).}

\end{figure}
One may ask whether the \citet{2002MNRAS_329_897H} formalism should
be believed. If $\dot{M}_{1}=-\dot{M}_{2}$, i.e. mass transfer is
conservative, one would expect $\dot{J}_{\mathrm{orb}}=0$ which is
clearly not the case. This has lead us to consider a simpler prescription
in which material lost from the binary system carries away some multiple
of the specific orbital angular momentum,

\begin{eqnarray}
\dot{J}_{\mathrm{orb}} & = & l\times\frac{J_{\mathrm{orb}}}{M_{1}+M_{2}}\left(\dot{M}_{1}+\dot{M}_{2}\right)\,,\label{eq:my-jorbdot}\end{eqnarray}
where $l$ is a free parameter. This formula has the advantage that
$\dot{J}_{\mathrm{orb}}=0$ in the case of conservative mass transfer.
It is difficult to constrain $l$ but Fig.~\ref{fig:ba-elogP-jorb-l-formalism}
shows the case $l=2$ which looks very similar to Fig.~\ref{fig:ba-elogP-jorb}
but without the requirement of a very large $\alpha_{\mathrm{W}}$.
The long-period tail is always present, but the number of stars we
predict there is small.

\section{Discussion}

\label{sec:discussion} The physical mechanism for the white dwarf
kick is unknown but candidates include magnetic fields \citep{1998A&A...333..603S},
asymmetric mass loss \citep{2003ApJ...595L..53F} and mass transfer
at periastron \citep{2000A&A...357..557S}. According to the evidence
from globular clusters the mechanism must be at work in both single
and binary stars \citep{2008MNRAS.383L..20D}.

We are working on an alternative mechanism which may be of relevance
to barium stars, that of eccentricity pumping due to a circumbinary
disk (Dermine~et~al.\emph{ in~prep.}), but again this cannot be
active in single stars. It would seem that asymmetric AGB wind loss
and/or interaction with a magnetic field is the leading candidate
if white-dwarf kicks are to occur in \emph{both} single and binary
stars. Still, a number of open questions remain and we briefly address
a few of them here.

\subsection{Fast and slow kicks -- and in which direction?}

In all our models presented above we have assumed an instantaneous
kick is given to the white dwarf at the end of the primary AGB phase.
However, any kick which is imparted by an asymmetric wind is likely
to be effective during the superwind phase of the primary AGB star,
i.e. on a timescale of $\sim10^{4}\,\mathrm{years}$. This is of the
same order as the tidal circularisation timescale and much longer
than the orbital period. A coupling of the orbital dynamics with a
{}``slow'' kick is beyond the scope of this paper but there may
still be an effect on the eccentricity particularly if the kick is
coupled to the orbital phase.

The direction of the kick is also an issue crucial to the further
evolution of the binary system. Close barium stars, with $P\lesssim\Pcirc\,\mathrm{days}$,
are circular by the end of the AGB so any kick increases the eccentricity
of the orbit. In wide systems which remain eccentric throughout the
AGB phase the eccentricity may increase or decrease depending on the
kick direction. For simplicity, we choose a fixed kick speed and a
random direction: this may not be realistic in a binary system in
which there is a clear axis of symmetry. The presence or otherwise
of a preferred direction would depend on the kick mechanism, which
is still unclear.

\subsection{Disrupted systems}

\label{sub:disruption}%
\begin{figure}
\includegraphics[bb=50bp 100bp 554bp 770bp,angle=270,scale=0.4]{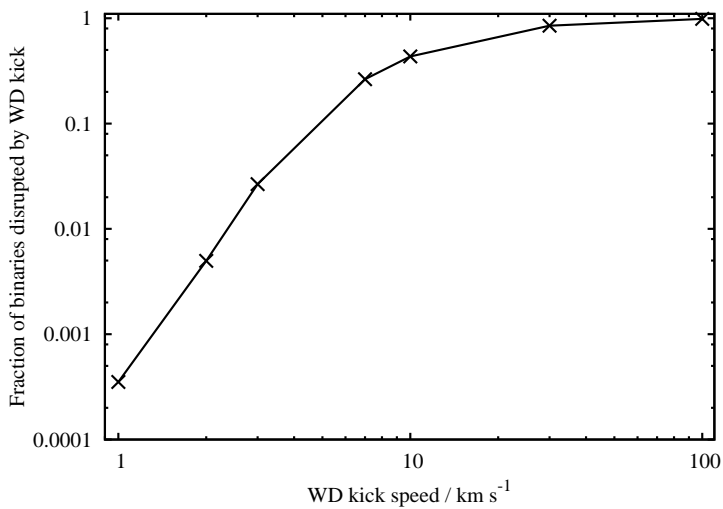}

\caption{\label{fig:disrupt}The fraction of binary systems which are disrupted
as a result of a white-dwarf kick as a function of the kick speed.
The models are as those of Section~\ref{sub:with-kicks} ($\lambda_{\mathrm{ion}}=0$,
$\alpha_{\mathrm{W}}=1.5$ and orbital angular momentum loss according
to Eq.~\ref{eq:hurley-jorbdot}).}

\end{figure}

A consequence of kicking a component of a binary star which is well
known in the study of massive stars is disruption of the binary system.
The fraction of binaries which are broken as a result of kicks in
our otherwise canonical model (c.f. Section~\ref{fig:ba-elogP-with-kicks})
is shown in Fig.~\ref{fig:disrupt}. For small velocities, $\lesssim3\,\mathrm{km}\,\mathrm{s}^{-1}$,
the number of broken systems is less than a few per cent. This is
a useful constraint on the kick mechanism because \citet{1998A&A...332..877J}
present 65 \emph{binary} barium stars, and possibly one single star
in a complete strong barium star sample, hence the disruption fraction
is less than one in 65, or about $1.5\%$, corresponding to a maximum
$\sigma_{\mathrm{WD}}=2-3\,\mathrm{km}\,\mathrm{s}^{-1}$. To the
authors' knowledge no conclusively-single \emph{giant }barium stars
have yet been found, but their number may be few and they may have
been missed. Some may be present among subgiant CH stars.

Wide binaries are most easily disrupted for a given $\sigma_{\mathrm{WD}}$.
We find a smaller breakup fraction when we consider our models with
efficient orbital angular momentum loss because these barium stars
have shorter periods and so are more tightly bound. For example, using
Eq.~\ref{eq:my-jorbdot} with $l=4$ we obtain a breakup fraction
of only $0.7\%$ with $\sigma_{\mathrm{WD}}=4\,\mathrm{km}\,\mathrm{s}^{-1}$
compared to $10\%$ in our canonical model with the same kick speed.

\subsection{Mass loss, the angular momentum budget and Mira}

\label{sub:angmom-budget}

\Change{We have hitherto ignored the uncertainty introduced by our
choice of mass loss rate. Our chosen rate, that of \citet{1993ApJ...413..641V},
tends to be small during most the TPAGB but strong during the final
superwind phase. This is advantageous for barium star formation as
it allows for many thermal pulses and associated barium production.
A smaller but more steady mass-loss rate may lead to fewer and weaker
third dredge up episodes because the envelope is more quickly reduced
in mass, but conversely any barium dredged up is less diluted so the
amount of barium in transferred material may be similar. In any case,
the efficiency of accretion is so poorly understood that the abundance
of barium in the primary is unlikely to be the largest uncertainty
in our analysis (although see \citealp{2007MNRAS.375.1280S,2009ApJ...696..797C}
for suitable analyses).}

It is not unreasonable to assume efficient loss of orbital angular
momentum, e.g. Eq.~\ref{eq:my-jorbdot} with $l>1$, if there is
significant locking of the stellar wind with a magnetic field or mass
is lost from a circumstellar or circumbinary disc. An alternative
explanation is that the \citet{2002MNRAS_329_897H} angular momentum
formula, our Eq.~\ref{eq:hurley-jorbdot}, is correct in the case
of efficient accretion on the companion. Efficient accretion may already
have been seen in systems such as Mira in which the stellar wind is
channelled onto the companion star \citep{2005ApJ...623L.137K}. Recent
simulations of wind-RLOF hybrid mass transfer suggest accretion is
indeed efficient \citep{2007ASPC..372..397M}. An alternative into
which we are looking is the implementation of the work of \citet{2009ApJ...702.1387S}
which would allow us to follow non-conservative mass transfer in detail,
but this is beyond the scope of this paper.

\subsection{Mild vs strong barium stars}

\label{sub:mild-strong}The differences between the distribution of
strong and mild barium stars in the $e-\log P$ diagram may help us
constrain uncertain physics. Our models predict that strong barium
stars are more likely to be made in the intermediate period range
$10^{3}-10^{4}\,\mathrm{days}$ because at short periods RLOF truncates
the primary AGB evolution while at long periods wind accretion is
not efficient. In contrast, mild barium stars are made over the whole
period range of $10^{2}-10^{5}\,\mathrm{days}$. The \citet{1998A&A...332..877J}
data may support this view. For the $30$ mild barium stars with known
periods, the mean period is $2900\,\mathrm{days}$ with a standard
deviation of $1900\,\mathrm{days}$, while for the $35$ strong barium
stars with known periods the mean period is $2000\,\mathrm{days}$
with a standard deviation of $1700\,\mathrm{days}$. The new \emph{HERMES}
data further lengthens the mean period of the mild barium stars.

There are five strong (binary) barium stars with periods less than
$500\,\mathrm{days}$ which prompt questions about their origin. It
is possible that the presence of a companion star enhances third dredge
up through the effects of tidal locking and rotational mixing (the
stellar core will not spin at the same rate as the envelope, we are
working on detailed models of AGB binaries to test this hypothesis).
Also, mass accretion may occur during the common envelope phase and
lead to strong\emph{ }barium star formation \citep{2008ApJ...672L..41R}.

In principle we could use the properties and number of barium stars
to constrain AGB physics, such as the amount of third dredge up, the
efficiency of the $^{13}\mathrm{C}$ pocket and the amount of $s$-process
element production. In practice this is difficult because we can increase
the amount of third dredge up in but decrease the $^{13}\mathrm{C}$
pocket efficiency in our models yet obtain the same barium/GK giant
fraction. Similarly, if we artificially increase the amount of dredge
up in the primary AGB star we can reduce the efficiency of accretion
on the secondary and obtain the same result. For these reasons we
have chosen the standard $^{13}\mathrm{C}$ pocket of \citet{2001ApJ...557..802B}
with the efficiency of third dredge up as predicted by \citet{Parameterising_3DUP_Karakas_Lattanzio_Pols}.
At a metallicity $Z=0.008$ we find that the Ba/GK-giant number ratio
is $\sim1\%$, as observed. 

Similarly, we could perhaps use the masses of barium stars \citep[e.g.][]{2006A&A...454..895A}
to constrain the amount of accreted material, but without prior knowledge
of the abundance of the accreted material, the amount of (or lack
of) thermohaline mixing or the initial mass of the (now) barium star,
this is difficult if not impossible, for example see \citet{2009PASA...26..176H}
and the CEMP star equivalent \citet{2009PASA...26..314B}.

\subsection{Implications for planetary nebulae}

\label{sub:PNe}Stars leaving the AGB pass through a planetary nebula
phase before they go on to become white dwarfs. Any asymmetry in the
AGB superwind might manifest itself as an asymmetric planetary nebula
and indeed most planetary nebulae have an axisymmetric rather than
spherical structure. A large number present large-scale deviations
from axisymmetry which can be explained by the presence of a binary
companion. 

\citet{1998ApJ...496..842S} investigated how an eccentric binary
can explain deviations from axisymmetry and \citet{2000A&A...357..557S}
proposed that the origin of the eccentricity is enhanced mass loss
at periastron passages. However, this process requires a seed eccentricity
which is not present in short-period barium-star progenitor systems
which are circularised by tides.

As white-dwarf kicks apparently occur in both single and binary stars,
we should see the effect of the kick in single-star planetary nebulae.
However, it is difficult to infer a kick from the structure of a planetary
nebula, e.g. a shift in the position of the central star relative
to the nebula, because its distortion can arise from the interaction
between the nebula and the interstellar medium or the proper motion
of the central star.

\section{Conclusions}

\label{sec:conclusions}The combination of a white-dwarf kick with
a speed of a few $\mathrm{km}\,\mathrm{s}^{-1}$, efficient common-envelope
ejection and orbital angular momentum loss enables us to much better
match the distribution of eccentricities and periods of observed barium
stars with our models. A small kick speed is compatible with observations
of both globular clusters and barium stars. Efficient common-envelope
ejection is required to prevent barium-star systems from spiralling
to short periods during the common envelope phase. 

Our orbital angular momentum loss mechanism is less certain. With
the existing formalism in our binary model, based on \citet{2002MNRAS_329_897H},
we can shrink barium-star orbits into the observed period range with
an accretion efficiency of $80\%$. However, we are probably extrapolating
their wind accretion model beyond its validity. If, instead, we assume
that material lost from the system carries away twice the specific
orbital angular momentum we achieve a similar result.

In any case, a white-dwarf kick is a plausible mechanism for generating
the eccentricity observed in the barium stars. It is supported by
independent evidence from globular clusters and is apparently not
a uniquely binary-star phenomenon. If it acts in barium stars it should
also be of consequence to the formation of CH and CEMP stars. \Change{Hopefully,
future studies of asymmetric planetary nebulae, circumbinary disks
in (post-)AGB stars, and globular-cluster and field white dwarfs may
pin down both the origin of the white-dwarf kicks and barium-star
eccentricity. While white-dwarf kicks may explain the barium star
mystery, the evidence as it stands is circumstantial and concrete
proof will only come from a combination of observations and improved
modelling.}
\begin{acknowledgements}
\Change{We would like to thank the referee, Oscar Straniero, for
his very helpful comments.} RGI would like to thank Sophie Van Eck,
Lionel Siess, Melvyn Davies and Chris Tout for many useful discussions
and in particular Alain Jorissen and Thomas Masseron for both useful
discussions and proofreading, John Fregeau for his inspirational talk
on white-dwarf kicks at the Lorentz Centre workshop \emph{Stellar
Mergers} and Lund Observatory for lending him the visitors' office.
RGI is the recipient of a \emph{Marie Curie-Intra European Fellowship}
at ULB. \Change{The research leading to these results has received
funding from the Seventh Framework Programme of the European Community
under grant agreement 220440.} TD is \emph{Boursier F.R.I.A.} \emph{
}RPC would like to thank the \emph{Wenner-Gren foundation} for a stipend.
\end{acknowledgements}

\end{document}